\documentclass[a4paper,12pt]{article}

\usepackage{amsmath}
\usepackage{amssymb}
\usepackage[dvips]{graphicx}
\usepackage[dvips]{psfrag}
\usepackage{bigdelim,multirow}
\usepackage{cite}

\makeatletter
\@addtoreset{equation}{section}
\renewcommand{\theequation}{\thesection.\@arabic\c@equation}
\makeatother

\newcommand{\deriv}[2]{\dfrac{\partial #1}{\partial #2}}

\newcommand{\dd}{{\rm d}}
\newcommand{\ee}{{\rm e}}

\begin{document}

\titlepage

\vspace*{-15mm}   
\baselineskip 10pt   
\begin{flushright}   
\begin{tabular}{r}   
{\tt APCTP-Pre2009-006}\\
{\tt arXiv:0906.2267}\\
June 2009 
\end{tabular}   
\end{flushright}   
\baselineskip 24pt   
\vglue 10mm   

\begin{center}
{\Large\bf
 Conformal Symmetry for Rotating D-branes
}

\vspace{8mm}   

\baselineskip 18pt   

\renewcommand{\thefootnote}{\fnsymbol{footnote}}

Li-Ming~Cao\footnote[2]{caolm@apctp.org}, 
Yoshinori~Matsuo\footnote[3]{ymatsuo@apctp.org}, 
Takuya~Tsukioka\footnote[4]{tsukioka@apctp.org} 
and 
Chul-Moon~Yoo\footnote[5]{c\_m\_yoo@apctp.org}

\renewcommand{\thefootnote}{\arabic{footnote}}
 
\vspace{5mm}   

{\it  
 Asia Pacific Center for Theoretical Physics, 
 Pohang, Gyeongbuk 790-784, Korea 
}
  
\vspace{10mm}   

\end{center}

\begin{abstract}
We apply the Kerr/CFT correspondence to 
the rotating black $p$-brane solutions. 
These solutions give the simplest examples 
from string theory point of view. 
Their near horizon geometries have structures of AdS,  
even though black $p$-brane solutions do not have 
AdS-like structures in the non-rotating case. 
The microscopic entropy which can be calculated via 
the Cardy formula exactly agrees with Bekenstein-Hawking 
entropy. 
\end{abstract}

\baselineskip 18pt   

\newpage

\section{Introduction}\label{sec:introduction}

Recently, the Bekenstein-Hawking entropy of Kerr black hole 
is calculated from the asymptotic symmetry~\cite{ghss}. 
In this calculation, the Cardy formula is 
used to relate the central charge of algebra to the entropy. 
This fact implies that there exists a 
correspondence between the Kerr black hole and 
some conformal field theory (CFT). 
The correspondence obtained is realized through a parallel way of  
the work by Brown and Henneaux~\cite{bh_1986} for AdS$_3$.%
\footnote{
There are also some works in which 
the entropy is calculated in a different formulation 
\cite{Carlip:1998wz,Park:2001zn,Kang:2004js}. 
} 
The authors in~\cite{ghss} used the near horizon extremal Kerr 
geometry which was found by Bardeen and Horowitz~\cite{bh_1999} instead. 

After the work~\cite{ghss}, the analysis was applied to 
various rotating black 
holes~\cite{Loran:2008mm,hhknt,lmp,hmns,cclp,pw,cw,ls,cmn,as,gg,kk,acott,g}.
However, it is difficult to obtain more information 
about the corresponding CFT than their symmetries. 
There are also some works on the geometries 
which can be related to the string 
theory~\cite{aot_0811,n,itw,aos,g_0901,lmpv,h}, 
but the explicit structures of their CFT sides are still unclear. 

In this paper, we apply the analysis of 
the Kerr/CFT correspondence 
to the rotating black $p$-brane solutions. 
These solutions are expected to describe 
the geometry in the presence of rotating D-branes. 
The corresponding field theory might be 
obtained from the low energy effective theory of D-branes, 
and will give one of the simplest examples 
from the viewpoint of string theory. 

This paper is organized as follows: 
In section~\ref{sec:solution}, 
we show the general form of the rotating black $p$-brane solutions, 
and their structure in the near horizon limit. 
A briefly review on the asymptotic symmetry and 
the formulation of the central charge is given in 
section~\ref{sec:symmetry}. 
We then derive the asymptotic symmetry of 
the rotating black $p$-brane solutions and 
calculate their central charges in section~\ref{sec:algebra}. 
In section~\ref{sec:correspondence}, 
we show the correspondence between 
the Bekenstein-Hawking entropy and 
the microscopic entropy which is obtained from the Cardy formula. 
Section~\ref{sec:conclusion} is devoted to 
conclusions and discussions.

\section{Rotating black $p$-brane solutions}\label{sec:solution}

We consider the rotating black $p$-brane solutions
in type II supergravity.
The D$p$-brane has a charge associated
with RR $(p+1)$-form field $A_{p+1}(x)$.
Relevant parts of the action are
\begin{equation}
 S
  =
  \frac{1}{2\kappa^2}
  \!\int\!\dd^{10}x \sqrt{-g}
  \left[
   \ee^{-2\Phi}
   \left( R + 4 (\partial\Phi)^2\right)
   -\frac{1}{2} \left|F_{p+2}\right|^2
  \right],
\end{equation}
where $\Phi(x)$ is the dilaton and
$F_{p+2}(x)=\dd A_{p+1}(x)$ is the field strength of
the RR $(p+1)$-form field.
The rotating black $p$-brane solution is given 
in~\cite{Cvetic:1996ek,Cvetic:1999ne,cdhlllmpst} (see also~\cite{ho}). 
The metric has the form of
\begin{align}
 \dd s^2
 &=
 H^{-1/2}
 \left[
  -H \dd t^2 + \dd\vec x^2
  + \frac{2m}{r^{D-3}}f_D
  \left(
   \cosh\delta \dd t + \sum_i a_i \mu_i^2 \dd\phi_i
  \right)^2
 \right]
 \notag\\&\quad
 + H^{1/2}
 \left[
  \frac{\dd r^2}{f_D\left(\Pi_D-\frac{2m}{r^{D-3}}\right)}
  + \sum_{i=1}^{\left[\frac{10-p}{2}\right]}
  \left(
   r^2 + a_i^2
  \right)
  \left(
   \dd\mu_i^2 + \mu_i^2 \dd\phi_i^2
  \right)
 \right], 
\end{align}
where $D=10-p$. For even $D$, we have to take 
$a_{i=(10-p)/2}=0$ and $\dd \phi_{i=(10-p)/2}=0$, 
since there are $\frac{10-p}{2}$ $\mu$'s but 
number of $\phi_i$ is $\frac{8-p}{2}$. 
Angular coordinates $\mu_i$ are not independent and
satisfy a constraint $\sum_i \mu_i^2 = 1$.
The functions $H(r, \mu)$, $f_D(r, \mu)$ and $\Pi_D(r)$ are
defined as 
\begin{align}
 H
 &=
 1 + f_D \frac{2m\sinh^2\delta}{r^{D-3}} ,
 &
 f_D^{-1}
 &=
 \Pi_D \left[\sum_i\frac{\mu_i^2}{g_i}\right] ,
 &
 \Pi_D
 &=
 \prod_i g_i ,
 &
 g_i
 &=
 1 + \frac{a_i^2}{r^2}.
\nonumber
\end{align}
The RR $(p+1)$-form field $A_{p+1}(x)$ and the dilaton $\Phi(x)$ are
given by 
\begin{align}
 A_{p+1}
  &=
  \frac{1}{\sinh\delta}
  \left(
   H^{-1} -1
  \right)
  \left(
   \cosh\delta \dd t + \sum_i a_i \mu_i^2 \dd\phi_i
  \right)
  \wedge \dd x^1 \wedge \cdots \wedge \dd x^p,
&
\\
 \ee^\Phi
  &=
 g_s H^\frac{3-p}{4}.
&
\end{align}

We now consider the extremal case.
The extremal limit can be realized through
the degeneracy of inner and outer horizons.
This degeneracy requires the following
condition at the horizon radius $r_H$, 
\begin{align}
\Pi_D\bigl|_{r=r_H} - \frac{2m}{r_H^{D-3}}= 0,
\nonumber
 \\
\Pi_D'\bigl|_{r=r_H} + (D-3)\frac{2m}{r_H^{D-2}} = 0,
\nonumber
\end{align}
where the prime (`` $'$ '') stands for the derivative
with respect to $r$.

Next we shall consider the near horizon geometry.
The near horizon metric has an AdS like structure 
in the components of $g_{tt}(x)$ and $g_{rr}(x)$ with 
$g_{t\phi_i}(x)$~\cite{bh_1999}. 
In order to see this, 
we need to consider the following coordinate transformations and
the limit, 
$$
r\to r_H+\epsilon\hat{r},
\qquad
t\to
 \epsilon^{-1} \hat t,
 \qquad
 \phi_i\to\varphi_i-\frac{a_i}{r_H^2 + a_i^2}\frac{1}{\cosh\delta}\
 \epsilon^{-1} \hat t
\qquad
\mbox{with}
\qquad
\epsilon\to 0.
$$
The coordinate transformation of $\phi_i$ can be easily
fixed by the condition of $g_{t\phi_i}(x)\sim\mathcal O(\epsilon^0)$.
Then this transformation makes all inverse powers of $\epsilon$ vanish.
Finally, we arrive at the near horizon geometry of 
the rotating black $p$-brane solution:  
\begin{align}
 \dd s^2
 &=
 f_0(\mu) \hat r^2 \dd\hat t^2
 + \gamma_{ij}(\mu)
 \left(\dd\varphi_i + k_i \hat r \dd\hat t\right)
 \left(\dd\varphi_j + k_j \hat r \dd\hat t\right)
 \notag\\&\quad
 + c_r^2 f_0(\mu) \frac{\dd\hat r^2}{\hat r^2}
 + \sum_i f_{\mu i}(\mu) \dd\mu_i^2
+ f_x(\mu)\dd\vec x^2,
\label{NearHorizonMetric}
\end{align}
where
\begin{align*}
 f_0(\mu)
 &=
 -\frac{k_0 \hat H^{1/2}}{2\hat F_D\cosh^2\delta}, 
 \\
 f_{\mu i}(\mu)
 &=
 \hat H^{1/2} \left(r_H^2 + a_i^2\right), 
\\
 f_x(\mu)
 &=
 \hat H^{-1/2},
\\
 \gamma_{ij}(\mu)
&=
  \hat H^{-1/2} \hat F_D a_i\mu_i^2 a_j \mu_j^2
  + \delta_{ij} \hat H^{1/2} \left(r_H^2+a_i^2\right)\mu_i^2 ,  
\end{align*}
with 
\begin{align*}
 \hat H(\mu)
&= H\Bigr|_{r=r_H} \!\!\!= 1 + \hat F_D \sinh^2\delta,
\\
\hat F_D(\mu)
&= f_D \Pi_D \Bigr|_{r=r_H}\!\!\!
 = \frac{2m}{r^{D-3}}f_D\Bigr|_{r=r_H}\!\!\!
 = \left(\sum_i\frac{r_H^2 \mu_i^2}{r_H^2 + a_i^2}\right)^{-1}.
\end{align*}
Constants $c_r$ and $k_i$ are given by 
\begin{align*}
  c_r 
 &= \frac{2\cosh\delta}{k_0},
 &
  k_0
 &=
 \left.\frac{\Pi_D''-\left(\frac{2m}{r^{D-3}}\right)''}
 {\Pi_D}\right|_{r=r_H}\!\!\!\!\!\! ,  &
 k_i
 &=
 - \frac{2 r_H a_i}{(r_H^2 + a_i^2)^2}\frac{1}{\cosh\delta} . 
\end{align*}
The RR field should be also transformed as
\begin{align*}
 A_{p+1}
 &=
 \epsilon^{-1}\tanh\delta\
 \dd \hat t \wedge \dd x^1 \wedge \cdots \wedge \dd x^p
 + \hat A_{p+1},
\end{align*}
where
\begin{align}
 \hat A_{p+1}
 &=
 \frac{\hat F_D}{\hat H} \sinh\delta
 \left[
  \sum_i a_i \mu_i^2 \left(\dd\varphi_i + k_i \hat r \dd\hat t\right)
 \right]
 \wedge \dd x^1 \wedge \cdots \wedge \dd x^p.
\label{RR}
\end{align}
The first term of $A_{p+1}$ is constant and 
can be absorbed by the gauge transformation. 
Then only the second term $\hat A_{p+1}$  
should be take into account in the near horizon limit. 
Hereafter, we will consider this near horizon
metric (\ref{NearHorizonMetric}) and the RR field (\ref{RR}) and omit
the hat (`` $\hat{\left.\right.}$ '').

\section{Asymptotic symmetry}\label{sec:symmetry}

In this section, we briefly review on 
the covariant formalism of the asymptotic symmetry~\cite{bb, bc}. 

Let us consider a system with local fields $\phi^i(x)$  
on $n$-dimensional spacetime. 
The variation of Lagrangian density can be 
divided into the ``left hand side'' of 
the equation of motion (EOM) and surface term: 
\begin{equation}
 \delta_I (\ast L) = (\text{EOM})_i\delta_I\phi^i + \dd \Theta_I, 
\label{variation_l}
\end{equation}
where $\delta_I$ is the variation with respect to the fields, 
$\phi^i(x)\to \phi^i(x) + \Delta\varphi^I\delta_I\phi^i(x)$,  
and $\Theta_I(x)$ is an $(n-1)$-form. 
The equation of motion is just given by (EOM)$_i=0$. 
If one restricts the configuration space to that of on-shell, 
the first term in (\ref{variation_l}) (EOM)$_i$ vanishes. 
The conserved current $J_\xi$ is given by the variation of the Lagrangian 
with respect to a symmetry:  
\begin{equation}
 \dd \tilde J_\xi = -(\text{EOM})_i\delta_\xi\phi^i ,   
\end{equation}
where, $\tilde J_\xi = \ast J_\xi$. 
If fields $\phi^i(x)$ satisfy the equations of motion, 
the current satisfies a condition $\dd\tilde J_\xi = 0$. 
Except for singularities, the current is expressed 
in terms of exact forms: 
\begin{equation}
 \tilde J_\xi = \dd k_\xi, 
\end{equation}
where $k_\xi$ is an arbitrary $(n-2)$-form.  
The charge is given by integrating this $(n-2)$-form, 
\begin{equation}
 Q_\xi = \!\int_{\Sigma}\! \tilde J_\xi 
= \!\int_{\partial\Sigma}\! k_\xi[\phi] . 
\end{equation}
Since $k_\xi$ is arbitrary, this charge cannot be determined without 
fixing reference.   

For asymptotic symmetry, we allow small deviations of the fields 
up to some asymptotic conditions\footnote{
The ${\cal O}(\chi^i)$ are symbolic forms to specify the asymptotic 
conditions. 
Later we will write this as ${\cal O}(r^\alpha)$ with some power
$\alpha$,  
where $r$ is a radial coordinate. 
}: 
\begin{equation}
  \phi^i \to \bar \phi^i + \mathcal O(\chi^i),  
\end{equation}
where $\bar{\phi}^i(x)$ are background fields.  
The current conservation is now  
satisfied up to some asymptotic condition,  
\begin{equation}
 \dd\tilde J_\xi = \mathcal O(\chi), 
\end{equation}
and the current is fixed up to the boundary term and 
the asymptotic condition, 
\begin{equation}
 \tilde J_\xi= \dd k_\xi + \mathcal O(\chi^\mu). 
\label{AsymptCurrent}
\end{equation}
In order to fix the $(n-2)$-form $k_\xi$, 
it might be useful to introduce the homotopy operator 
which is defined such that 
\begin{align}
 \delta \omega_n &= \delta\phi^i E_i + \dd I^n \omega_n, \label{HomotopyN}
 \\
 \delta \omega_p &= I^{p+1}\dd\omega_p + \dd I^p \omega_p,  \label{HomotopyP}
\end{align}
where $\omega_p$ is an arbitrary $p$-form. 
The homotopy operator 
$I^p$ maps $p$-form to $(p-1)$-form,  
\begin{align}
 I^n \omega_n 
 &= 
 \delta\phi\,\deriv{(i_{\partial_\mu}\omega_n)}{\phi_{,\mu}} 
 + \delta\phi_{,\nu}\deriv{(i_{\partial_\mu}\omega_n)}{\phi_{,(\mu\nu)}} 
 - \delta\phi\,\partial_\nu\deriv{(i_{\partial_\mu}\omega_n)}{\phi_{,(\mu\nu)}} 
 + \cdots ,
 \\
 I^{n-1} \omega_{n-1} 
 &= 
 \frac{1}{2} \delta\phi\,\deriv{(i_{\partial_\mu}\omega_{n-1})}{\phi_{,\mu}} 
 + \frac{2}{3} \delta\phi_{,\nu}
 \deriv{(i_{\partial_\mu}\omega_{n-1})}{\phi_{,(\mu\nu)}} 
 - \frac{1}{3} \delta\phi\,\partial_\nu
 \deriv{(i_{\partial_\mu}\omega_{n-1})}{\phi_{,(\mu\nu)}} 
 + \cdots , \label{ExplicitP}
\end{align}
where $\phi_{,\mu\cdots\nu}(x)\equiv\partial_\mu\cdots\partial_\nu\phi(x)$. 
One can define an asymptotic charge which could be observed 
through the difference from the background $\bar{\phi}^i(x)$.
This definition corresponds to $J_\xi[\bar\phi]=0$, and 
therefore we choose $k_\xi[\bar\phi]=0$ 
by using the ambiguity of $k_\xi$. 
Then the current of asymptotic symmetry can be expanded as 
\begin{equation}
 J^\mu_\xi[\phi] = J^\mu_\xi[\bar\phi] 
  + \delta J^\mu_\xi[\delta\phi,\bar\phi] + \cdots
  = \delta J^\mu_\xi[\delta\phi,\bar\phi] 
  + \mathcal O(\chi^\mu), \label{CurrentExpand}
\end{equation}
where $\delta\phi^i(x) = \phi^i(x) - \bar\phi^i(x)$.  
From the equations \eqref{AsymptCurrent}, \eqref{HomotopyP} 
and \eqref{CurrentExpand}, 
we obtain the following expression: 
\begin{equation}
 k_\xi[\phi-\bar\phi,\bar\phi] 
  = I^{n-1} \tilde J_\xi + \mathcal O(\chi^{\mu\nu}). \label{DefinitionK}
\end{equation}
The asymptotic charge is then given by using this $k_\xi$ 
(\ref{DefinitionK}), 
\begin{equation}
 Q_\xi = \!\int_{\partial\Sigma}\! k_\xi[\delta\phi, \bar{\phi}] . 
\label{charge_00}
\end{equation}

The commutation relation of the asymptotic charges can be calculated by 
taking the variation of the field $\phi^i(x)$: 
\begin{equation}
 \delta_\xi Q_\zeta =\!\int\! k_\zeta[\delta_\xi\phi,\bar\phi].  
\end{equation}
Since the charges are linear order in 
$\delta\phi^i(x) = \phi^i(x)-\bar\phi^i(x)$, 
this can be expressed as 
\begin{equation}
 \delta_\xi Q_\zeta = \!\int\! k_\zeta[\delta_\xi\bar\phi,\bar\phi] 
  + \!\int\! k_\zeta[\delta_\xi\delta\phi,\bar\phi]. 
\label{GeneralCommutator}
\end{equation}
The second term which is linear in $\delta\phi^i(x)$ gives 
the commutator of $\xi(x)$ and $\zeta(x)$.
An additional constant term appears in the first term, 
which gives the central charge of this algebra. 

Before closing this section, it would be worth mentioning  
a symplectic structure of the configuration space~\cite{lw, w, iw}.   
We start by introducing the symplectic form
\begin{equation}
 \Omega = \frac{1}{2}\Omega_{IJ} \dd\varphi^I \wedge \dd\varphi^J,  
\end{equation}
where $\dd\varphi^I$ are the basis of one-form on the configuration space 
and $\Omega_{IJ}$ is given by 
\begin{equation}
 \Omega_{IJ} = \!\int\! \omega_{IJ}, 
\end{equation}
with
\begin{equation}
 \omega_{IJ} = \delta_I\Theta_J - \delta_J\Theta_I . 
\end{equation}
If one add the boundary term $K$ to the Lagrangian, 
the symplectic form gets an additional term, $[\delta_I,\delta_J]K$. 
Therefore this definition of the symplectic form  
depends on choice of the boundary term. 
By definition, a function $H_\xi$ defines a vector field 
$V^I_\xi(\phi^i)$ as 
\begin{equation}
 \delta_J H_\xi = \Omega_{IJ} V^I_\xi,  
\label{hamiltonian}
\end{equation}
therefore $H_\xi$ is the charge which generates 
the flow in the configuration space, 
\begin{equation}
 \delta_\xi \phi^i = V^I_\xi \delta_I \phi^i . 
\end{equation} 

In order to relate this symplectic structure to 
the asymptotic charge, one could introduce an additional term 
and use the following definition of the symplectic form instead:  
\begin{equation}
 \omega_{IJ} = \delta_I\Theta_J - \delta_J\Theta_I - \dd E_{IJ}, 
\end{equation}
with
\begin{equation}
 E_{IJ} = 
 -\frac{1}{2}\left(I^n_I\Theta_J - I^n_J\Theta_I\right). 
\end{equation}
Then, it was shown in~\cite{bb,bc} that 
the symplectic form is related to the $(n-2)$-form $k_\xi$ 
which is defined in \eqref{DefinitionK} as 
\begin{equation}
 \dd k_\xi[\delta_J\phi,\bar\phi] 
= I^n_J \dd \tilde J_\xi - \delta_J \tilde J_\xi 
= \omega_{IJ}V^I_\xi , 
\label{dk_xi}
\end{equation}
if the variation $\delta_J$ is tangent to the space of solutions. 
The charge $H_\xi$ which is defined in \eqref{hamiltonian} 
is identical to the one in (\ref{charge_00}).

\section{Virasoro algebra}\label{sec:algebra}

We now study the asymptotic symmetry of 
the rotating black $p$-brane solution. 
Here we generalize the analysis in~\cite{ghss}, 
and consider a geometry of the following form 
which might be extended from the one in (\ref{NearHorizonMetric}), 
\begin{align}
 \dd s^2
 &=
 f_0(y) r^2 \dd t^2
 + \gamma_{ij}(y)
 \left(\dd\varphi_i + k_i r \dd t\right)
 \left(\dd\varphi_j + k_j r \dd t\right)
 \notag\\&\quad
 + c_r^2f_0(y) \frac{\dd r^2}{r^2}
 + \sigma_{ab}(y) \dd x^a \dd x^b 
 + \tau_{\alpha\beta}(y) \dd y^\alpha \dd y^\beta , 
 \label{GeneralMetric}
\end{align}
where $i,j=1,\cdots,d$, $a,b=1,\cdots,\bar d$ and 
$\alpha,\beta=1,\cdots,\tilde d$. 
We impose the following asymptotic boundary condition 
for the deviation of the metric $h_{\mu\nu}(x)$: 
\begin{equation}
h_{\mu\nu}
=
\bordermatrix{
  & t & i & k \neq i & r & b & \beta \cr
t & {\cal O}(r^2) & {\cal O}(r^0) 
& {\cal O}(r^1) & {\cal O}(r^{-2}) & {\cal O}(r^{-1}) & {\cal O}(r^{-1}) 
\cr
i & & {\cal O}(r^0) & {\cal O}(r^0) & {\cal O}(r^{-1}) 
& {\cal O}(r^{-1}) & {\cal O}(r^{-1}) \cr
j \neq i & & & {\cal O}(r^{-1}) & {\cal O}(r^{-1}) & {\cal O}(r^{-1})
& {\cal O}(r^{-1}) \cr
r & & & & {\cal O}(r^{-3}) & {\cal O}(r^{-2}) & {\cal O}(r^{-2}) \cr
a & & & & & {\cal O}(r^{-1}) & {\cal O}(r^{-1}) \cr
\alpha & & & & & & {\cal O}(r^{-1}) \cr
}
\label{boundary_condition}
\end{equation}
where the index ``$i$'' stands for a specific direction 
of the Killing $\varphi_i$, and ``$j$'' and ``$k$'' are 
other directions of $\varphi$. 
Indices of $a$, $b$, $\alpha$, $\beta$ indicate 
$x^a$, $x^b$, $y^\alpha$, $y^\beta$ direction, respectively. 

We can also introduce a $(p+1)$-form gauge field of the form 
\begin{equation}
 A_{p+1} = \sum_{i} A_i(y)\left(\dd\varphi_i + k_i r\,\dd t\right) 
  \wedge \dd x^1 \wedge \cdots \wedge \dd x^p . 
\end{equation}
In~\cite{cmn}, an asymptotic condition is imposed on 
the gauge field and a part of the transformation 
is absorbed by using the $U(1)$ gauge transformation. 
However we do not impose any condition on the gauge field here. 
Since constraints on the geometry are enough to 
fix the asymptotic Killing vectors, 
we do not need to impose any more constraints on the gauge field. 
Since asymptotic symmetry is defined up to some asymptotic condition, 
the theory could be symmetric even though it is not invariant. 
In fact, the gauge field does not contribute to the central charge 
even if we do not impose asymptotic condition. 
As we will see later, these contributions can be absorbed 
by redefinition of the asymptotic charge. 
Therefore we do not need to absorb the transformation 
of gauge fields by using $U(1)$ gauge transformation. 

We introduce a dilaton $\Phi(y)$. 
Due to the isometry of the geometry, 
the dilaton does not depends on $t$, $\varphi$, and $x$. 
We also assume that the dilaton does not have any singularity 
or zero at the horizon of the original geometry. 
Then the dilaton depends on only $y^\alpha$ coordinates. 
Since it is rather natural to consider 
the D-brane effective theory in string frame, 
we consider the asymptotic charge and the Virasoro algebra 
in the string frame. 

The diffeomorphism which satisfies this asymptotic condition 
\eqref{boundary_condition} is given for each rotating directions 
\begin{equation}
 \xi^{(i)} 
  = \epsilon(\varphi_i) \partial_{\varphi_i} 
  - r \epsilon'(\varphi_i) \partial_r . 
\end{equation}
Using the mode expansion for $\epsilon(\varphi_i)$, 
we define $\xi^{(i)}_n$ as 
\begin{equation}
 \xi^{(i)} = \sum_n\epsilon_n\xi^{(i)}_n 
\qquad 
\text{with} 
\qquad 
 \xi^{(i)}_n 
 = e^{-in\varphi_i} \left( \partial_{\varphi_i} + inr\partial_r\right) . 
\end{equation}
These vectors satisfy the Virasoro algebras, 
\begin{equation}
 i [\xi^{(i)}_n,\xi^{(j)}_m] = \delta_{ij} (n-m) \xi^{(i)}_{n+m} . 
\end{equation}

Let us proceed to calculate the central charges. 
In our case, the dilaton is invariant under this diffeomorphism. 
Therefore contributions to the central charges come from 
the conserved $(n-2)$-form associated to 
the gravity and the RR $(p+1)$-form. 
The (on-shell vanishing) current is thus given by 
\begin{equation}
 J^\mu_\xi = J^\mu_{\xi(g)} + J^\mu_{\xi(A)}, 
\end{equation}
where $J^\mu_{\xi(g)}$ and $J^\mu_{\xi(A)}$ are 
the currents associated to the equations of motion of 
gravity and RR $(p+1)$-form, respectively. 
Their explicit forms are 
\begin{align}
 J^\mu_{\xi(g)} 
 &= \ee^{-2\Phi}\biggl[R^{\mu\nu} - \frac{1}{2} g^{\mu\nu} R 
 + 2 \nabla^\mu \nabla^\nu \Phi  - 2 g^{\mu\nu} \nabla_\rho \nabla^\rho \Phi 
 \notag\\&\qquad\qquad
 + 2 g^{\mu\nu} \nabla_\rho \Phi \nabla^\rho \Phi \biggr] \xi_\nu
  - T^{\mu\nu} \xi_\nu, \\
 J^\mu_{\xi(A)} &= (p+2) 
  (\nabla_\nu F^{\nu\mu\lambda_1\cdots\lambda_p}) 
 \xi^\rho A_{\rho\lambda_1\cdots\lambda_p}, 
\end{align}
where $\nabla_\mu$ is the covariant derivative. 
An energy-momentum tensor for the RR field is expressed by $T^{\mu\nu}(x)$.  
We introduce a small perturbation of the metric and the RR $(p+1)$-form, 
so that the conserved $(n-2)$-forms $k_\xi$ are given by 
\eqref{DefinitionK}. 
We define $\tilde k_\xi$ as the dual of $k_\xi$, 
which can be derived from the current $J_\xi$ 
by using \eqref{ExplicitP} and \eqref{DefinitionK}. 
The relevant formula is given by 
\begin{equation}
 \tilde k_\xi^{\mu\nu}[\delta\phi]
 = 
 \frac{1}{2} \delta\phi^I\,
 \deriv{J^\mu_\xi}{\phi^I_{,\nu}} 
 + \frac{2}{3} \delta\phi^I_{,\lambda}
 \deriv{J^\mu_\xi}{\phi^I_{,(\nu\lambda)}} 
 - \frac{1}{3} \delta\phi^I\,\partial_\lambda
 \deriv{J^\mu_\xi}{\phi^I_{,(\nu\lambda)}} 
 - (\mu\leftrightarrow\nu) . 
\end{equation}
The gravity part of $\tilde k_\xi$  
with the metric perturbation $\delta g_{\mu\nu}(x) = h_{\mu\nu}(x)$ is 
given by 
\begin{align}
 \tilde k^{\mu\nu}_{\xi(g)}[h] 
 = - \frac{1}{2\kappa^2} \ \ee^{-2\Phi}
 \Bigl[&
  \left(D^\nu h\right)\xi^\mu
  + \left(D_\sigma h^{\mu\sigma}\right)\xi^\nu
  + \left(D^\mu h^{\nu\sigma}\right)\xi_\sigma 
 \notag\\&
  + \frac{1}{2} h D^\nu \xi^\mu
  + \frac{1}{2} h^{\mu\sigma} D_\sigma \xi^\nu
  + \frac{1}{2} h^{\nu\sigma} D^\mu \xi_\sigma
 \notag\\&
  + \frac{3}{2} \xi^\mu h^{\nu\rho} \partial_\rho\Phi
  + \frac{1}{2} h^{\mu\rho}\xi_\rho \partial^\nu\Phi 
  - \frac{1}{2} \xi^\mu h \partial^\nu\Phi
 \Bigr]
 - (\mu\leftrightarrow\nu),  
\end{align}
where $D_\mu$ is the covariant derivative for the 
background $\bar g_{\mu\nu}(x)$, 
while the RR field part with metric perturbation is expressed by  
\begin{align}
 \tilde k^{\mu\nu}_{\xi(A)}[h]
 = - \frac{p+2}{2\kappa^2} 
 \Bigl[&
  2 F^{\nu\rho\lambda_1\cdots\lambda_p}h^\mu_\rho
  \xi^\sigma A_{\sigma\lambda_1\cdots\lambda_p}
  + p F^{\nu\mu\rho\lambda_1\cdots\lambda_{p-1}}h^{\kappa}_\rho
  \xi^\sigma A_{\sigma\kappa\lambda_1\cdots\lambda_{p-1}}
 \notag\\&
  - \frac{1}{2} F^{\nu\mu\lambda_1\cdots\lambda_p} 
  h \xi^\sigma A_{\sigma\lambda_1\cdots\lambda_p}
 \Bigr]
 - (\mu\leftrightarrow\nu) . 
\end{align}
The gravity part with the gauge 
perturbation $\delta A_{\mu_1\cdots\mu_{p+1}}(x)
=a_{\mu_1\cdots\mu_{p+1}}(x)$ is 
\begin{align}
 \tilde k^{\mu\nu}_{\xi(g)}[a] 
 = \frac{p+2}{2\kappa^2} 
 \biggl[
  & (p+1) \xi_\sigma F^{\sigma\nu\rho_1\cdots\rho_p} 
  a^\mu_{\ \rho_1\cdots\rho_p} 
  + (p+1) F^{\mu\nu\rho_1\cdots\rho_p} \xi^\sigma 
  a_{\sigma\rho_1\cdots\rho_p} 
 \notag\\&
  + 2 \xi^{\nu} F^{\mu\rho_1\cdots\rho_{p+1}} 
  a_{\rho_1\cdots\rho_{p+1}}
 \biggr] 
 - (\mu\leftrightarrow\nu) , 
\end{align}
and the RR $(p+1)$ form part is given by 
\begin{align}
 \tilde k^{\mu\nu}_{\xi(A)}[a] 
 = \frac{p+2}{2\kappa^2} 
 \biggl[
  & \left( D^\nu a^{\mu\lambda_1\cdots\lambda_p} \right) 
  \xi^\sigma A_{\sigma\lambda_1\cdots\lambda_p} 
  - \frac{1}{2} a^{\mu\lambda_1\cdots\lambda_p} 
  D^\nu \left(\xi^\sigma A_{\sigma\lambda_1\cdots\lambda_p}\right) 
 \notag\\&
  + \frac{p}{3} \left( D^\rho a^{\nu\mu\lambda_1\cdots\lambda_{p-1}} \right) 
  \xi^\sigma A_{\sigma\rho\lambda_1\cdots\lambda_{p-1}}
\notag\\
&
  + \frac{p}{3} \left( D_\rho a^{\rho\mu\lambda_1\cdots\lambda_{p-1}} \right) 
  \xi_\sigma A^{\sigma\nu}_{\ \ \lambda_1\cdots\lambda_{p-1}} 
 \notag\\&
  - \frac{p}{6} a^{\nu\mu\lambda_1\cdots\lambda_{p-1}} 
  D^\rho \left(\xi^\sigma A_{\sigma\rho\lambda_1\cdots\lambda_{p-1}}\right)
\notag
\\
&  - \frac{p}{6} a^{\rho\mu\lambda_1\cdots\lambda_{p-1}} 
  D_\rho \left(\xi_\sigma A^{\sigma\nu}_{\ \ \lambda_1\cdots\lambda_{p-1}}\right)
 \biggr]
 - (\mu\leftrightarrow\nu) . 
\end{align}

The central extensions could be calculated by 
the first term in l.h.s. of \eqref{GeneralCommutator}, 
\begin{equation}
 \frac{c^{(i)}}{12}\equiv \int_{\partial\Sigma}\! 
  k_{\xi^{(i)}_n}[\delta_{\xi^{(i)}_m}\bar\phi,\bar\phi],  
\end{equation}
where fields $\phi^i(x)$ are the metric and the RR $(p+1)$-form. 
We set the variations as their Lie derivatives of the background 
fields  
$h_{\mu\nu}(x)=\pounds_{\xi}\bar{g}_{\mu\nu}(x)$ and 
$a_{\mu_1\cdots\mu_{p+1}}(x)=\pounds_{\xi}A_{\mu_1\cdots\mu_{p+1}}(x)$.  
Explicit forms of the Lie derivatives for the metric are written down as
\begin{align}
 \pounds_{\xi^{(i)}_n} \bar{g}_{IJ} 
 &= 
 in e^{-in\varphi_i} (\delta_{It}+\delta_{Jt}-\delta_{Ii}-\delta_{Ji}) 
\bar{g}_{IJ},
 \\
 \pounds_{\xi^{(i)}_n} \bar{g}_{ir} 
 &= \pounds_{\xi^{(i)}_n} \bar{g}_{ri} 
 = n^2 r e^{-in\varphi_i} \bar{g}_{rr},  
 \\
 \pounds_{\xi^{(i)}_n} \bar{g}_{\mu\nu} &= 0, 
\qquad\text{(for other components)}
\end{align}
where indices $I$ and $J$ stand for $(t,\varphi_i, \varphi_j)$, 
and for the RR $(p+1)$-form as   
\begin{equation}
 \pounds_{\xi_n^{(i)}} A 
  = in \ee^{-in\varphi_i} 
  \sum_j A_j(y) \left(k_j r \dd t - \delta_{ij}\dd\varphi_i\right) 
  \wedge \dd x^1 \wedge \cdots \wedge \dd x^p . 
\end{equation}
Since two vectors $\xi^{(i)}_n$ and $\xi^{(i)}_m$ 
are manifestly antisymmetric, 
the central charge has only the odd power of $n$. 
In this case, we can easily see that 
it contains only terms of order $n$ and $n^3$. 
The linear part can be absorbed by 
taking an appropriate definition of $Q_0$. 
Hence we consider only the terms of order $n^3$. 
Using the fact that $i_{\xi_n} A = \mathcal O(n^0)\times e^{-in\varphi}$ 
and $\pounds_{\xi_n} A = \mathcal O(n) \times e^{-in\varphi}$, 
we see that there are no contributions to $\mathcal O(n^3)$ terms 
from $k_{\xi(g)}[\pounds_\zeta A]$, $k_{\xi(A)}[\pounds_\zeta A]$ 
and $k_{\xi(A)}[\pounds_\zeta \bar{g}]$. 
As a result, the remaining relevant part 
is $k_{\xi(g)}[\pounds_\zeta \bar{g}]$ 
and can be rewritten as 
\begin{align}
 k^{\mu\nu}_{\xi}[\pounds_{\zeta} \bar{g}]
 = 
 \frac{1}{2\kappa^2} \ \ee^{-2\Phi}
 \Bigl[&
  -D_\rho\xi^\rho D^\nu \zeta^\mu 
  +D_\rho\zeta^\rho D^\nu \xi^\mu 
  +2 D_\rho\xi^\nu D^\rho \zeta^\mu \notag\\
  &+ \frac{1}{2} 
    \left(
     D^\rho\xi^\nu + D^\nu \xi^\rho 
    \right)
    \left(
     D_\rho\zeta^\mu + D^\mu \zeta_\rho 
    \right)
  - 2 R^{\mu\rho} \xi^\nu \zeta_\rho 
  + R^{\mu\nu\rho\sigma}\xi_\rho\zeta_\sigma 
 \notag\\&
  - \frac{1}{2} \xi^\mu \left(D^\nu\zeta^\rho+D^\rho\zeta^\nu\right) 
 \partial_\rho\Phi 
  - \frac{3}{2} \left(D^\mu\zeta^\rho+D^\rho\zeta^\mu\right) 
 \xi_\rho \partial^\nu\Phi 
 \Bigr] 
 \notag\\
 - &(\mu\leftrightarrow\nu) , 
\end{align}
up to exact forms.  
By using the explicit expressions of the 
Lie derivatives with respect to the vector field $\xi^{i}_n$, 
it turns out that the terms to contribute to the central charges
are only  
\begin{equation}
\frac{c_i}{12}
=  
\frac{1}{2\kappa^2}
  \!\int\!\dd^d\varphi\,\dd^{\bar d}x\,\dd^{\tilde d}y 
\sqrt{-g} \ \ee^{-2\Phi}
  \Big[D_\rho\xi^{(i)t}_{n} D^\rho\xi^{(i)r}_m - (m\leftrightarrow n)
 \Big]. 
\end{equation}
Using \eqref{GeneralMetric}, we finally obtain the following
expressions,  
\begin{equation}
 \frac{c_i}{12} 
  = \frac{c_r k_i}{\kappa^2}
  \!\int\!\dd^d\varphi\,\dd^{\bar d}x\,\dd^{\tilde d}y  
  \ \ee^{-2\Phi}
  \sqrt{(\det \gamma)(\det\sigma)(\det\tau)}. 
  \label{CentralCharge}
\end{equation}

\section{Correspondence of the entropy}\label{sec:correspondence}

In order to see the correspondence, 
we have to define the quantum vacuum 
in the rotating black $p$-brane background. 
In the case of rotating black hole background, 
we have to measure the energy from the viewpoint of 
the zero angular momentum observer (ZAMO) 
due to the ergoregion of the geometry. 
Let us consider, for example, 
a field with an energy $\omega$ and angular momenta $m_i$. 
This field has an energy
\begin{equation}
 \omega_{\text{ZAMO}} = \omega - \Omega_i m_i , 
\end{equation}
where $\Omega_i$ is the angular velocity at the horizon. 
Then the Boltman factor for this field is given by 
\begin{equation}
 e^{-\beta_H \omega_{\text{ZAMO}}} 
  = e^{-\beta_H \omega + \beta_i m_i}, 
\end{equation}
where $\beta_H = 1/T_H$ is the inverse of the Hawking temperature. 
The Frolov-Thorne temperature $T_i = 1/\beta_i$~\cite{ft} is defined by 
\begin{equation}
 \beta_i = \beta_H\Omega_i . 
\end{equation}
This field could have a nontrivial Boltzmann factor 
even if the Hawking temperature is zero. 

In order to calculate the Frolov-Thorne temperature, 
we first consider the near-extremal case, 
and then take the extremal limit. 
The near extremal metric has the form of 
\begin{equation}
 \dd s^2 = - r^2\left(1-\frac{r_+^2}{r^2}\right) f_0 \dd t^2 
  + \gamma_{ij}\bigl(\dd\varphi_i + k_i r \dd t \bigr)
  \bigl(\dd\varphi_j + k_j r \dd t\bigl) 
  + \frac{f_r \dd r^2}{r^2\left(1-\frac{r_+^2}{r^2}\right)} + \cdots , 
\end{equation}
where $r_+$ is the position of the horizon. 
Since this is the near horizon geometry of the rotating black $p$-brane, 
this horizon corresponds to the outer horizon of original metric and 
the inner horizon is placed at $r=0$. 
The temperature and the velocity are given by 
\begin{align}
 T_H &= \frac{r_+}{2\pi}\sqrt{\frac{f_0}{f_r}} , &
 \Omega_i &= r_+ k_i, 
\end{align}
so that the Frolov-Thorne temperature can be estimated as  
\begin{equation}
 T_i = \frac{T_H}{\Omega_i} 
  = \frac{1}{\pi c_r k_i}. 
\end{equation}

By using the Cardy formula, 
the microscopic entropy is given by 
\begin{equation}
 S = \frac{\pi^2}{3}c_i T_i 
  = \frac{4\pi}{\kappa^2}\!\int\! \dd^d\varphi\,\dd^{\bar d}x\,\dd^{\tilde d}y 
  \ \ee^{-2\Phi}
  \sqrt{(\det \gamma)(\det\sigma)(\det\tau)} , 
\end{equation}
for each rotating directions $i$. 
The additional factor $\ee^{-2\Phi}$ 
is absorbed by the redefinition of the metric 
in the Einstein frame. 
This entropy exactly agrees with the Bekenstein-Hawking entropy. 

\section{Conclusions and discussions}\label{sec:conclusion}

In this paper, we applied the Kerr/CFT correspondence 
to the rotating black $p$-brane solutions. 
Near horizon geometries of these solutions are 
given by AdS$_2$ with $U(1)$ fibers. 
By imposing appropriate boundary conditions, 
we obtained the asymptotic symmetry which 
forms the Virasoro algebra. 
Assuming that there exists a corresponding CFT, 
we applied the Cardy formula 
and obtained the entropy from the central charge of the algebra. 
This microscopic entropy exactly agrees with 
the Bekenstein-Hawking entropy. 

These solutions are expected to be 
describe the geometry in presence of D$p$-brane. 
Hence, if there exist corresponding CFTs, 
they might be related to the low energy effective theory of D-branes. 
For the D0-brane, 
the effective theory is expected to be a 
one-dimensional field theory. 
This implies that the corresponding field theories
of rotating black holes are not two-dimensional theories 
but one-dimensional field theories 
(or spacetime is compactified to one dimension). 
In other words, 
the effective theory of D0-brane might be 
a one-dimensional conformal field theory 
for a special configuration 
which corresponds to the rotating case. 
Then, the conformal symmetry should be 
that on the direction of worldline. 
Actually, time direction of the black-$p$-brane geometry 
is included into the angular coordinate of 
their near horizon geometry. 
The RR field basically lies on the time direction, 
but on the rotational direction 
at the origin of the near horizon geometry 
on which it couples to the black $p$-brane. 
Hence all of the Virasoro algebras associated with 
the rotational directions pick up 
the same degree of freedom in the field theory side. 
In fact, the Bekenstein-Hawking entropy is 
reproduced by each of these Virasoro algebras, 
not sum of these contributions. 
Therefore, the degree of freedom associated to these 
Virasoro symmetries should be identical to each other. 

It is interesting that the effective theories of 
rotating D-branes might be a CFT even though 
general non-rotating D-brane effective theories are not CFT. 
There exists a difference between the 
near horizon geometry of the rotating black $p$-brane 
and that of the non-rotating case. 
The non-rotating black $p$-brane solution does not 
have an AdS-like structure while 
we obtain the AdS$_2$ geometry in the rotating case. 
This implies that there should be some links  
between the rotating and non-rotating cases 
also in the effective theory of the D-branes. 
It is also interesting to consider 
how this conformal symmetry appears in the effective theories. 
This is left for future studies. 

\vspace*{5mm}

\noindent
 {\large{\bf Acknowledgments}}

This work is supported by YST program in APCTP. 

\vspace*{2mm}


\end{document}